\documentclass[aps, a4paper, superscriptaddress, nofootinbib,  twocolumn, 10pt]{revtex4-1}
\usepackage{epsfig}
\usepackage{graphicx}
\usepackage{slashed}
\usepackage{hyperref}
\usepackage{mathtools}

\newcommand{\beq}{\begin{equation}}
\newcommand{\eeq}{\end{equation}}
\newcommand{\bea}{\begin{eqnarray}}
\newcommand{\eea}{\end{eqnarray}}
\newcommand{\ba}{\begin{align}}
\newcommand{\ea}{\end{align}}
\newcommand{\bfig}{\begin{figure}}
\newcommand{\efig}{\end{figure}}

\newcommand{\D}{\displaystyle}

\newcommand{\gev}{\, \text{GeV}}
\newcommand{\mev}{\, \text{MeV}}

\newcommand{\tin}{t_{in}}

\usepackage{color}

\newcommand{\omnes}{{\cal{O}}}

\begin{document}

\phantom{}
\vspace*{-17mm}

\title{Test of analyticity and unitarity for the pion form-factor data around the $\rho$ resonance}
\author{B.Ananthanarayan}
\affiliation{Centre for High Energy Physics,
Indian Institute of Science, Bangalore 560 012, India}
\author{Irinel Caprini}
\affiliation{Horia Hulubei National Institute for Physics and Nuclear Engineering,
P.O.B. MG-6, 077125 Magurele, Romania}
\author{ Diganta Das}
\affiliation{Department of Physics and Astrophysics, University of Delhi, Delhi 110007, India}



\begin{abstract}  High-statistics data on the $e^+e^-\to \pi^+\pi^-$ cross section and the pion vector form factor have been obtained recently by several collaborations.  Unfortunately, there are some tensions between  different datasets,  especially the most precise ones, which have not been resolved so far.  Additional independent constraints on the data are therefore of interest.  We consider a  parametrization-free method of analytic extrapolation proposed recently, which is based on a mixed phase and modulus extremal problem and combines rigorous upper and lower bounds  with numerical simulations to account for the statistical distributions of the input and output values.  Spacelike data on the form factor and measurements of the modulus in the region  $(0.65-0.71) \gev$ are used as input. In  previous works, the formalism was applied for extrapolating the form factor to low energies. In the present work, we use it as a  stringent and model-independent test of consistency with analyticity and unitarity for the high-statistics data around the $\rho$ resonance. The study reveals some inconsistencies, in particular below the $\rho$ peak  the BABAR data are slightly higher than the band of extrapolated values, while above the $\rho$ peak all the data  are situated at the lower edge of the band. The implications of the results on the two-pion vacuuum polarization contribution to the anomalous magnetic moment of the muon are briefly discussed.
\end{abstract}
\maketitle
\section{Introduction}

The  electromagnetic form factor $F_\pi^V(t)$ of the pion is a fundamental quantity for strong-interaction physics, intensively studied   theoretically and experimentally for more than sixty years. The recent interest in its precise determination is mostly driven by the anomalous magnetic moment of the muon, $a_\mu=(g-2)_\mu/2$.
 There is at present a  disagreement between the experimental value of the muon anomaly  measured at  BNL  \cite{Bennett:2006fi} and its evaluation in the Standard Model (SM).  Two new experiments aim at reducing the experimental uncertainty by a factor of four: the
E989 experiment at Fermilab  \cite{Grange:2015fou}, which started running in 2018, and the E34 experiment at J-PARC, which plans
to start its first run in 2024 \cite{Abe:2019thb}.
 In parallel, there is a continuous effort for improving the accuracy of the theoretical calculation of $a_\mu$ in the SM (for a recent  review and earlier references see \cite{Aoyama:2020ynm}).

The hadronic vacuum polarization (HVP) contribution to the muon $g-2$ is the leading hadronic contribution, which  cannot be calculated using perturbative QCD and dominates the theoretical uncertainty. By analyticity and unitarity, the hadronic loops contribute to leading order (LO) as a dispersion integral over the cross section of the $e^+e^-$ annihilation into hadrons. This allows a data-driven evaluation of HVP using experimental data (the most recent determinations are reported in \cite{Davier:2019can, Keshavarzi:2019abf, Aoyama:2020ynm}). At low energies, below $\sqrt{s}= 1 \gev$, the  dispersive integral is dominated by the cross section of the $e^+e^-$ annihilation into two pions.  Several high-statistics $e^+e^-$ experiments \cite{Akhmetshin:2003zn}-\cite{Ablikim:2015orh} have been designed and operated recently in order to measure this quantity with increased precision. However, as noted in \cite{Davier:2019can, Keshavarzi:2019abf, Aoyama:2020ynm}, there are some tensions between the data of different experiments, particularly the most precise ones, BABAR \cite{Aubert:2009ad, Lees:2012cj} and KLOE \cite{Ambrosino:2008aa,  Ambrosino:2010bv, Babusci:2012rp}.

It is useful to keep in mind that to LO the cross section of the $e^+e^-\to\pi^+\pi^-$ process is expressed in terms of  the modulus squared of the pion vector form factor, $F_\pi^V(t)$.  This allows, in principle, to use additional theoretical and experimental information available on the form factor in order to improve the accuracy, especially in regions where the data are still poor.  The most powerful approach for the study of the form factor is dispersion theory, which exploits analyticity, unitarity and crossing symmetry. The form factor has been calculated also on the lattice, but the results did not reach the same accuracy until now,  and we shall not consider them.

The dispersion theory of the pion form factor traditionally exploits  Fermi-Watson theorem  \cite{Fermi:2008zz, Watson:1954uc}, which states that below the first inelastic threshold the phase of the form factor on the unitarity cut  is equal to the $P$-wave phase shift  of $\pi\pi$ elastic scattering. By the well-known Omn\`es representation, the form factor is reconstructed as an analytic function in the whole complex $t$-plane from its phase of the boundary.  The most recent dispersive analysis  \cite{Colangelo:2018mtw} is based on a parametrization involving an Omn\`es function multiplied by  factors which account for inelastic and isospin-breaking  effects. The  $P$-wave $\pi\pi$ phase shift used as input was obtained by solving Roy equations, which fully exploit analyticity, unitarity and crossing symmetry of pion-pion scattering amplitudes, and the free parameters of the model have been fixed by fitting data on the modulus of the form factor from $e^+e^-$ experiments.  

As noted in \cite{Colangelo:2018mtw}, in the combined fit of all data one can  observe the well-known tension between the  BABAR and KLOE data: the BABAR data lie systematically above the KLOE data. The fit follows the average of the two in accordance with their respective covariance matrices. In the Omn\`es representation used in \cite{Colangelo:2018mtw}, an assumption about the phase in the inelastic region,  where it is not known, has been adopted. Actually, the parametrization contains an additional factor, which is complex above the inelastic threshold and can modify the initial phase. However, the independence of the results on the  choice of the phase in the Omn\`es function is not formally proved. 

As  remarked for the first time in  \cite{Caprini:1999ws}, in order to avoid assumptions on the unknown phase  in the inelastic region, one can use instead the phase some information available in this region on the modulus. This leads to a mixed phase and modulus representation, which can be generalized by including also an arbitrary number of values of the form factor at points inside the analyticity domain. The price to be paid for the independence on the unknown phase above the inelastic threshold is that the formalism cannot predict definite values, but only upper and lower bounds on the form factor. The bounds can be calculated exactly with techniques of functional analysis in terms of the information used as input (for technical detail see \cite{Abbas:2010jc, Caprini:2019osi}).  
 
An important step forward was achieved in Refs. \cite{Ananthanarayan:2016mns,   Ananthanarayan:2017efc, Ananthanarayan:2018nyx, Ananthanarayan:2019zic}, where the rigorous bounds derived from analyticity have been merged with numerical simulations which properly take into account the statistical distribution of the input and  output values. This elaborate formalism has been applied for the determination of the pion charge radius \cite{Ananthanarayan:2017efc} and the modulus of the form factor at low energies \cite{Ananthanarayan:2016mns, Ananthanarayan:2018nyx,Ananthanarayan:2019zic}. In particular, a more accurate value of the HVP contribution to $a_\mu$ from  energies below 0.63 GeV has been obtained.  The input used in these works consisted from the most recent measurements of the form factor in the spacelike region and the data on the modulus measured in  the range of  $(0.65-0.71) \gev$, where the various experiments are in somewhat better mutual agreement than in other regions.

In principle, the method can be used to predict the form factor also at higher energies, above 0.71 GeV, up to  the first inelastic threshold. However, as will be clear below, the analyticity bounds  become gradually weaker at energies far from the input range, especially when the inelastic threshold is approached. Therefore, unlike the low-energy region where the predictions have been more precise than the experimental data, above the 
$\rho$-resonance region, where precise data by high statistics experiments are available,  we do not expect our determinations to compete with them in precision. However,  the formalism is still useful since it provides a stringent test of consistency of the form-factor data  with analyticity and unitarity. 

The purpose of the present work is to explore the outcome of this test for energies below and above the $\rho$ resonance. The paper is organized as follows: in the next section, we give a brief description of the formalism, expressing it as a parametrization-free test of consistency for the values of the form factor at various energies. In Sec. \ref{sec:test} we describe the theoretical and phenomenological information used as input and the methodology for extrapolating the form factor in the output region. In Sec. \ref{sec:res} we present our results, confronting the extrapolated values with the experimental data in the region $(0.72-0.9)\gev$ as the modulus input are taken from the region $(0.65-0.71)\gev$. A discussion of the results, in particular of their implication on the HVP contribution to the muon $g-2$, is given in  Sec. \ref{sec:conc}.  The paper has an Appendix, which contains an explicit proof of the independence of the results on the unknown phase of the form factor above the inelastic threshold.

\section{Parametrization-free analyticity and unitarity constraints}\label{sec:method}
The pion vector form factor $F_\pi^V(t)$ is an analytic function 
in the complex $t$ plane  cut along  the real range $t\ge t_+$, where $t_+= 4 m_\pi^2$. It satisfies the Schwarz reflection property $F_\pi^V(t^*)=(F_\pi^V(t))^*$ and the normalization $F_\pi^V(0)=1$. 

Along the cut the form factor is a complex function.
According to Fermi-Watson theorem  \cite{Fermi:2008zz, Watson:1954uc}, below the first inelastic threshold the phase of the form factor is equal to the $P$-wave phase shift $\delta_1^1(t)$ of the $\pi\pi$ elastic scattering.  Since this theorem is valid in the exact isospin limit, we must first  remove the main  isospin-violating effect, known to arise from $\omega-\rho$ and  $\phi-\rho$  interference. 

We shall follow the standard approach \cite{Leutwyler:2002hm, Hanhart:2012wi, Colangelo:2018mtw} to do this, by defining a purely $I=1$ function $F(t)$ as
\beq\label{eq:F}
F(t)=F_\pi^V(t)/F_{\omega+\phi}(t),
\eeq
where the function $F_{\omega+\phi}(t)$, specified in the next section, includes the $I=0$ contribution due to $\omega$ and $\phi$ resonances. Then Fermi-Watson theorem allows us to write
\beq\label{eq:watson}
\arg F(t+i\epsilon)=\delta_1^1(t),  \quad\quad  4 m_\pi^2 \le t \le \tin,
\eeq
where one can assume with a good approximation that the first inelastic threshold is  set up by the $\omega\pi$-production threshold, and take  $\sqrt{\tin}=m_\omega+m_\pi=0.917 \gev$.  In the elastic region, the phase shift $\delta_1^1(t)$ is known with high precision  from  dispersion theory for $\pi\pi$ scattering \cite{Ananthanarayan:2000ht, Caprini:2011ky, GarciaMartin:2011cn}.

Above $\tin$, where the phase of the form factor is not known, we use the information available on the modulus from experimental measurements and perturbative QCD. Specifically, we adopt a conservative condition written as
\beq\label{eq:L2}
 \frac{1}{\pi} \int_{\tin}^{\infty} |F_\pi^V(t)|^2  \frac{dt}{t} \leq  I
 \eeq
where the integral  converges and allows an accurate evaluation of $I$ from the available information. 

From the phase and modulus conditions (\ref{eq:watson}) and (\ref{eq:L2}),  using techniques of functional analysis and optimization theory (for a review see \cite{Caprini:2019osi}), one can derive constraints on the values of the form factor and its derivatives  at points inside the analyticity domain. Omitting the proof given in \cite{Abbas:2010jc, Caprini:2019osi} (see also the Appendix of 
\cite{Ananthanarayan:2018nyx}), we shall write down these constraints in a compact form, by expressing the form factor  in the  isospin limit as
\beq\label{eq:gF}  F(t)= \frac{\omnes(t)}{\omega(t)}\, \frac{g(\tilde z(t))}{w(\tilde z(t))},
\eeq
where
\beq	\label{eq:omnes}
\omnes(t) = \exp \left(\D\frac {t} {\pi} \int^{\infty}_{t_+} dt' 
\D\frac{\phi(t^\prime)} {t^\prime (t^\prime -t)}\right),
\eeq
\beq\label{eq:omega}
\omega(t) =  \exp \left(\D\frac {\sqrt {\tin - t}} {\pi} \int^{\infty}_{\tin}  \D\frac {\ln |\omnes(t^\prime)|\, {\rm d}t^\prime}
{\sqrt {t^\prime - \tin} (t^\prime - t)} \right),
\eeq 
\beq\label{eq:outer}
w(z)=\sqrt{\frac{1-z}{1+z}},
\eeq 
and the function 
\beq\label{eq:ztin}
\tilde z(t) = \frac{\sqrt{\tin} - \sqrt {\tin -t}} {\sqrt{\tin} + \sqrt {\tin -t}}
\eeq
conformally maps the $t$ complex plane cut for $t\ge \tin$   onto the unit disk $|z|<1$, such that $\tilde z(0)=0$ and the upper(lower) edges of the cut become the upper(lower) semicircles in the $z$ plane.

We recall that $\omnes(t)$ is an Omn\`es function, where
 $\phi(t)$ is equal to $\delta_1^1(t)$ for
$t\le \tin$ and is an arbitrary function above $\tin$.  The function denoted as $\omega(t)$ is analytic in the  $t$ plane cut for $t>\tin$
and has the modulus equal to  $|\omnes(t)|$ on the cut, and $w(z)$ is an outer function (analytic and without zeros in the unit disk $|z|<1$), with modulus on the boundary $|z|=1$ of the disk related to to the weight in the integral (\ref{eq:L2}).

Finally, the function $g(z)$ appearing in (\ref{eq:gF})  is an analytic function in the disk $|z|<1$ and satisfies the boundary condition  
\beq\label{eq:gI1}
 \frac{1}{2 \pi} \int^{2\pi}_{0} {\rm d} \theta |g(\zeta)|^2 \leq I, \qquad \zeta=e^{i\theta}.
 \eeq
This condition implies rigorous correlations 
among the values of the function  $g(z)$ and its derivatives 
at points inside the holomorphy domain, $|z|<1$. In particular,
  if $z_n\in (-1, 1)$, $n=1,\,2,\,3$ are three arbitrary real points,  the following positivity condition holds 
\beq\label{eq:posit}
{\cal D}\ge 0
\eeq
 where ${\cal D}$ is the determinant  defined as
\beq\label{eq:det}
{\cal D}=\left|
\begin{array}{c c c c c c}
I-g(0)^2 & \xi_{1} & \xi_{2} &   \xi_{3}\\
	\xi_{1} & \D \frac{z^{2}_{1}}{1-z^{2}_1} & \D
\frac{z_1z_2}{1-z_1z_2}  &  \D  \frac{z_1z_3}{1-z_1z_3} \\
	\xi_{2} & \D \frac{z_1 z_2}{1-z_1 z_2} & 
 \D  \frac{z_2^2}{1-z_2^2}  &  \D  \frac{z_2z_3}{1-z_2z_3} \\
	\xi_3 & \D \frac{z_1 z_3}{1-z_1 z_3} & \D
\frac{z_2 z_3}{1-z_2 z_3} &  \D  \frac{z_3^2}{1-z_3^2} \\
	\end{array}\right|,
\eeq
 where
\beq\label{eq:g0}
g(0)=w(0) \omega(0) /\omnes(0),
\eeq
and
\beq\label{eq:barxi}
  \xi_n =g(z_n) - g(0), \quad 1\leq n \leq 3.\eeq
Moreover, the principal minors of the determinant ${\cal D}$ must also be  nonnegative. We emphasize that the normalization condition $F_\pi^V(0)=1$ is included as input and is implemented in the definition of $g(0)$ in (\ref{eq:g0}).

The inequality (\ref{eq:posit}) defines an allowed domain for the real values $g(z_n)$, which can be expressed in a straightforward way in terms of the values of the form factor at the points $t_n=\tilde t(z_n)$, where
\beq
\tilde t(z)= \tin\, \frac{4 z}{(1+z)^2}
\eeq
is the inverse of (\ref{eq:ztin}).

In particular, for $t_n<t_+$, when both $F(t_n)$ and $\omnes(t_n)$ are real, we have from (\ref{eq:gF}):
\beq\label{eq:gFn}
g(z_n)= F(t_n) w(z_n)\, \omega(t_n) /\omnes(t_n), 
\eeq
while for $t_+<t_n<\tin$, when $F(t_n)$ and $\omnes(t_n)$ are complex and have equal phases,
\beq\label{eq:gFn1} g(z_n) = |F(t_n)| w(z_n)\, \omega(t_n) \, /|\omnes(t_n)|,
\eeq
 where the modulus $|\omnes(t)|$ of the Omn\`es function is obtained from (\ref{eq:omnes}) by the principal value (PV) Cauchy integral
\beq	\label{eq:modomnes}
 |\omnes(t)| = \exp \left(\frac {t} {\pi} \text{\rm PV} \int^{\infty}_{t_+} dt' 
\D\frac{\phi (t^\prime)} {t^\prime (t^\prime -t)}\right).
\eeq

Using these relations, the   inequality (\ref{eq:det}) can be written as an explicit consistency test imposed by anayticity on the values $F(t_n)$  at three arbitrary points below $\tin$.  As proved in \cite{Abbas:2010jc, Caprini:2019osi}, the constraint exploits in an optimal way the input information. Moreover, although the function $\omnes(t)$  depends on the phase $\phi(t)$ for $t>\tin$, the  test is independent on it. For completeness, we give in the Appendix, along with the formal argument put forward in  \cite{Caprini:1999ws, Abbas:2010jc}, an explicit proof of this independence, presented here for the first time.

\section{Determination of $|F_\pi^V|$ in the $\rho$ region}\label{sec:test}
In our analysis, we have treated two of the values of $F(t)$ in  (\ref{eq:det}) as input, using experimental information on the form factor (or its modulus) at two points on the spacelike and timelike axes. For each input, the inequality (\ref{eq:posit}) becomes a quadratic constraint on the form factor at the third point, from which upper and lower bounds are exactly derived. The test will consist from the comparison of these bounds  with the experimental data available in the third region. The regions used as  input and output will be specified below.

\subsection{Inputs}\label{sec:input}
We have extracted the pion form factor from the cross section of the $e^+e^-\to\pi^+\pi^-$ process using standard correction factors (see for instance Appendix B of \cite{Ananthanarayan:2016mns}). We then converted the  input values of  $F_\pi^V(t)$ into the isospin-conserving function $F(t)$ using (\ref{eq:F}), solved the optimization problem  for this function and finally re-expressed the results in terms of  $|F_\pi^V(t)|$.  The function $F_{\omega+\phi}(t)$, which accounts for the isospin violation due to $\omega-\rho$ and $\phi-\rho$ mixing, has been taken of the form \cite{Leutwyler:2002hm, Hanhart:2012wi}:

\begin{align}\label{eq:rhoomega}
 F_{\omega+\phi}(t)= 1+\frac{\epsilon\, t}{t-(m_\omega-i\Gamma_\omega/2)^2 }+\frac{\epsilon^\prime\, t}{t-(m_\phi-i\Gamma_\phi/2)^2},
\end{align}
where  $\epsilon=-2\times 10^{-3}$, $\epsilon^\prime=5\times 10^{-3}$ and the masses and the decay widths are taken as  $m_\omega=781.86$, $m_\phi = 1019.46 \mev$, 
$\Gamma_\omega=8.5 \mev$ and $\Gamma_\phi=4.3 \mev$ \cite{Tanabashi:2018oca}.

We have taken the phase shift $\delta_1^1(t)$  entering the integral   (\ref{eq:omnes}) below $\tin$ from  \cite{Ananthanarayan:2000ht, Caprini:2011ky} (Bern phase) and \cite{GarciaMartin:2011cn} (Madrid phase). They have been  calculated with good precision using dispersion relations  for $\pi\pi$ scattering and are mutually consistent, with slightly larger uncertainties of the Bern phase  near $\tin$. Above $\tin$ we have taken for $\phi(t)$
an arbitrary expression. As we have already mentioned, the results are independent of this arbitrariness. The proof of this is provided in  Appendix \ref{sec:A}.

We have calculated the integral (\ref{eq:L2}) using the BABAR data \cite{Aubert:2009ad} from $\tin$ up to $\sqrt{t}=3 \gev$, smoothly continued with  a constant value for the modulus in the range $3 \gev \leq \sqrt{t} \leq 20 \gev$,  and  a decrease  $\sim 1/t$ at higher energies, as predicted by perturbative QCD \cite{Farrar:1979aw, Lepage:1979zb}. With this input we have obtained $I=0.578 \pm 0.022$, where the uncertainty is  due to the BABAR experimental errors. In the calculations we have used as input for $I$  the central value increased by the  error, which leads to the most conservative bounds due to a monotonicity property discussed in \cite{Ananthanarayan:2016mns}.

The input at interior points was taken from the most recent experimental measurements of  $F_\pi$ Collaboration at JLab \cite{Horn:2006tm, Huber:2008id} on the spacelike axis, 
\bea\label{eq:Huber}	
F_\pi^V(-1.60 \gev^2)= 0.243 \pm  0.012_{-0.008}^{+0.019}, \nonumber \\ 
F_\pi^V(-2.45 \gev^2)=  0.167 \pm 0.010_{-0.007}^{+0.013}, 
\eea
and the modulus measured by the $e^+e^-$ experiments SND (04), CMD2 (06), BABAR (09), KLOE (11), KLOE (13) and BESIII (15) \cite{Akhmetshin:2003zn}-\cite{Ablikim:2015orh} in the region $(0.65 - 0.71) \gev$ on the timelike axis. 

The choice of the input region $(0.65 - 0.71) \gev$ was motivated in our  previous works \cite{Ananthanarayan:2016mns, Ananthanarayan:2017efc,  Ananthanarayan:2018nyx}  by the fact that  here the  data have a better precision than at the lower energies where we extrapolated the form factor \footnote{In Ref.[25] we have studied in detail the issue of sensitivity to this particular choice and have demonstrated the robustness of the method.}. The number of input points in this region from each experiment is given in Table 1 of Ref. \cite{Ananthanarayan:2016mns}.  We adopt the same input region in the present work, since the determinations of different   experiments, especially the most precise ones, BABAR and KLOE, are  more consistent among them in this region than at higher energies, around the $\rho$ resonance (see for instance the comparisons in Fig. 13 of \cite{Aoyama:2020ynm}).

We recall that the hadronic decays of $\tau$ leptons have been used in the past as an alternative source of  data in the evaluation of the HVP contribution to $a_\mu$. Isospin-breaking (IB) and electromagnetic corrections must be applied in order to convert  these data to the cross section of $e^+e^-$ annihilation  (for a recent evaluation and earlier references see \cite{1808470}). However, as remarked in \cite{Aoyama:2020ynm},  the present understanding of these corrections is  not yet at a level allowing the use of $\tau$ data in the muon $g-2$ determinations.  In view of this consensus in the community, in this work we will use only $e^+e^-$ data.  

\begin{figure*}[thb]\vspace{0.1cm}
	\begin{center}
		\includegraphics[width = 6.1cm]{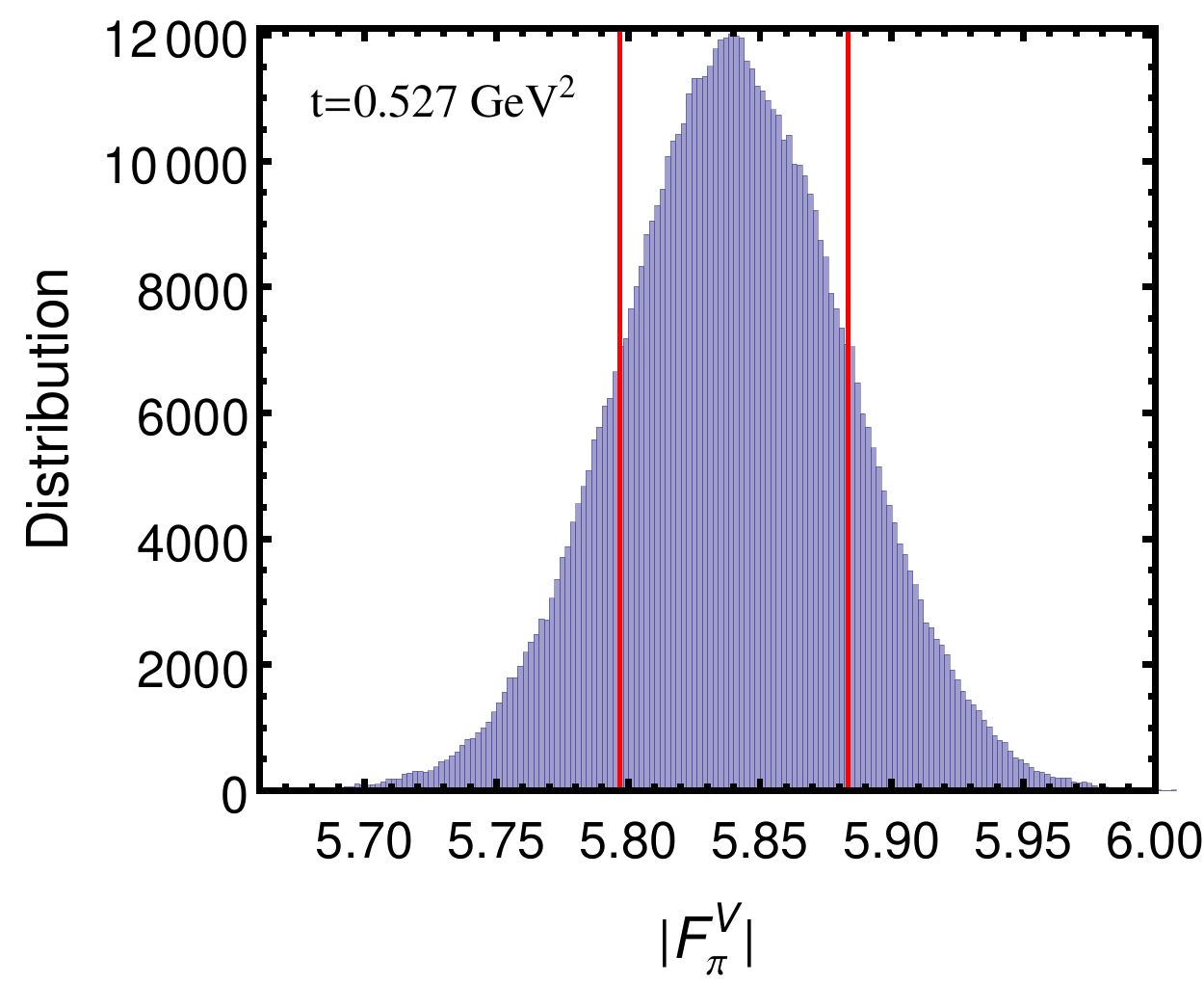}
		\includegraphics[width = 6cm]{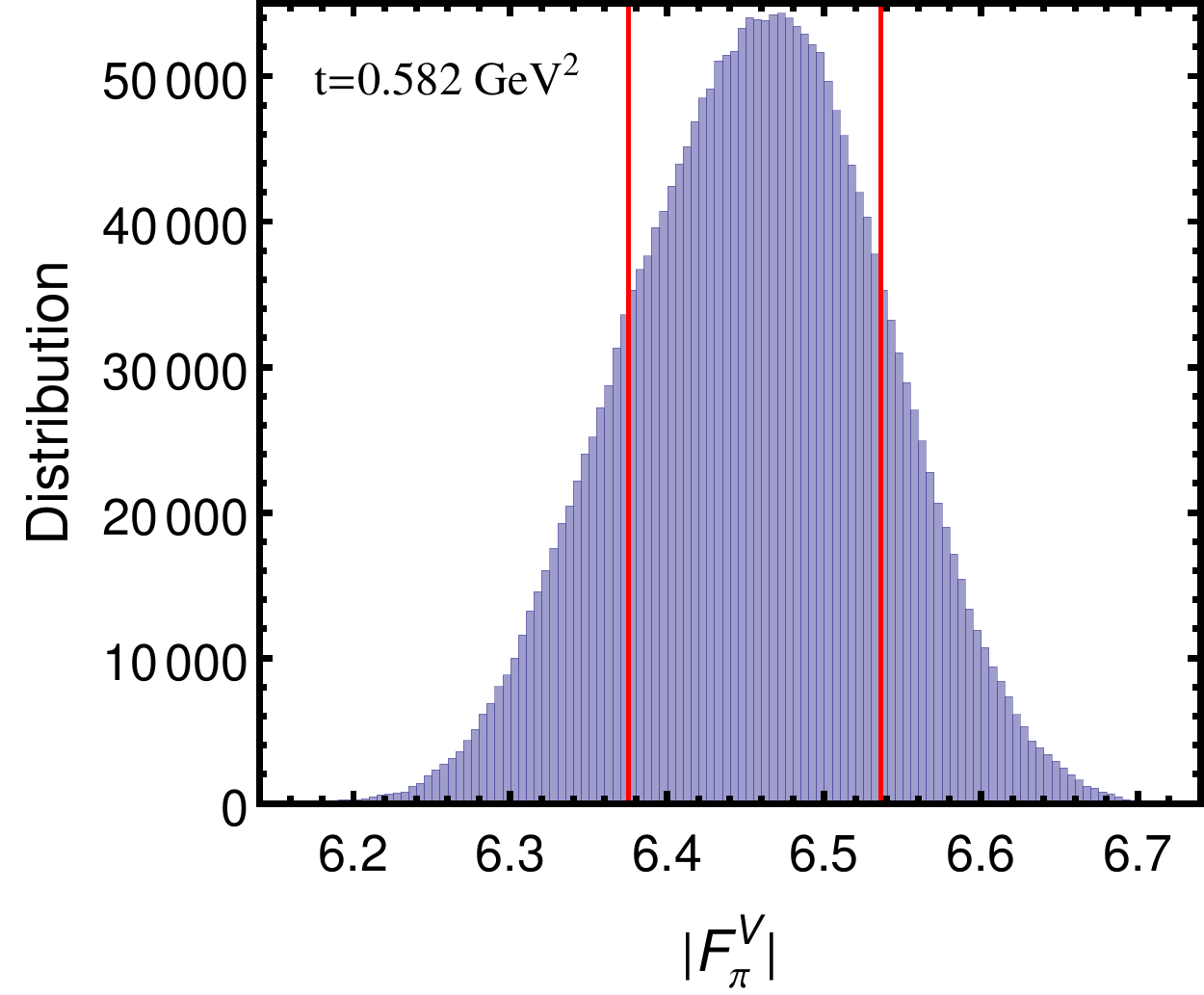}
		\includegraphics[width = 6cm]{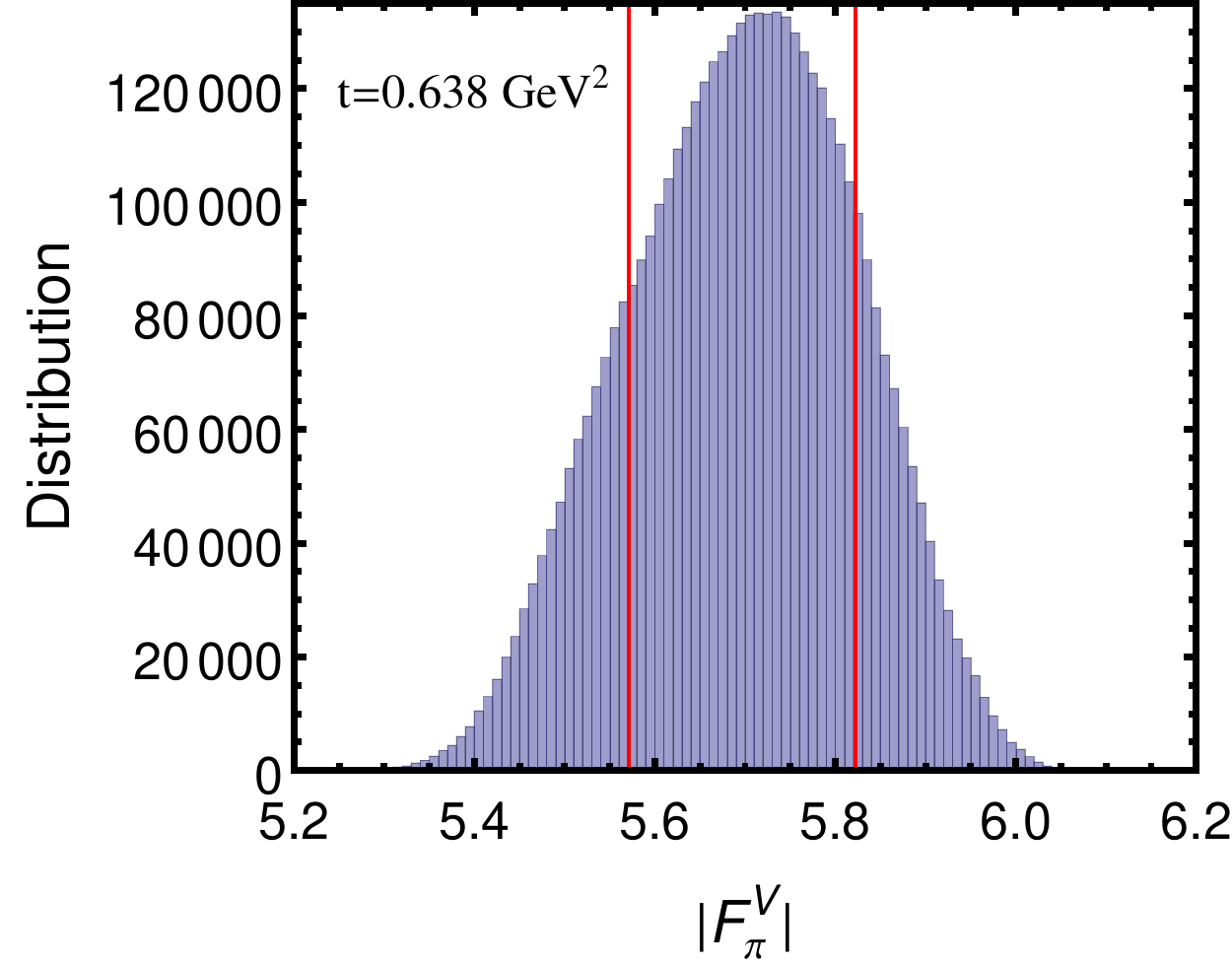}
		\includegraphics[width = 6cm]{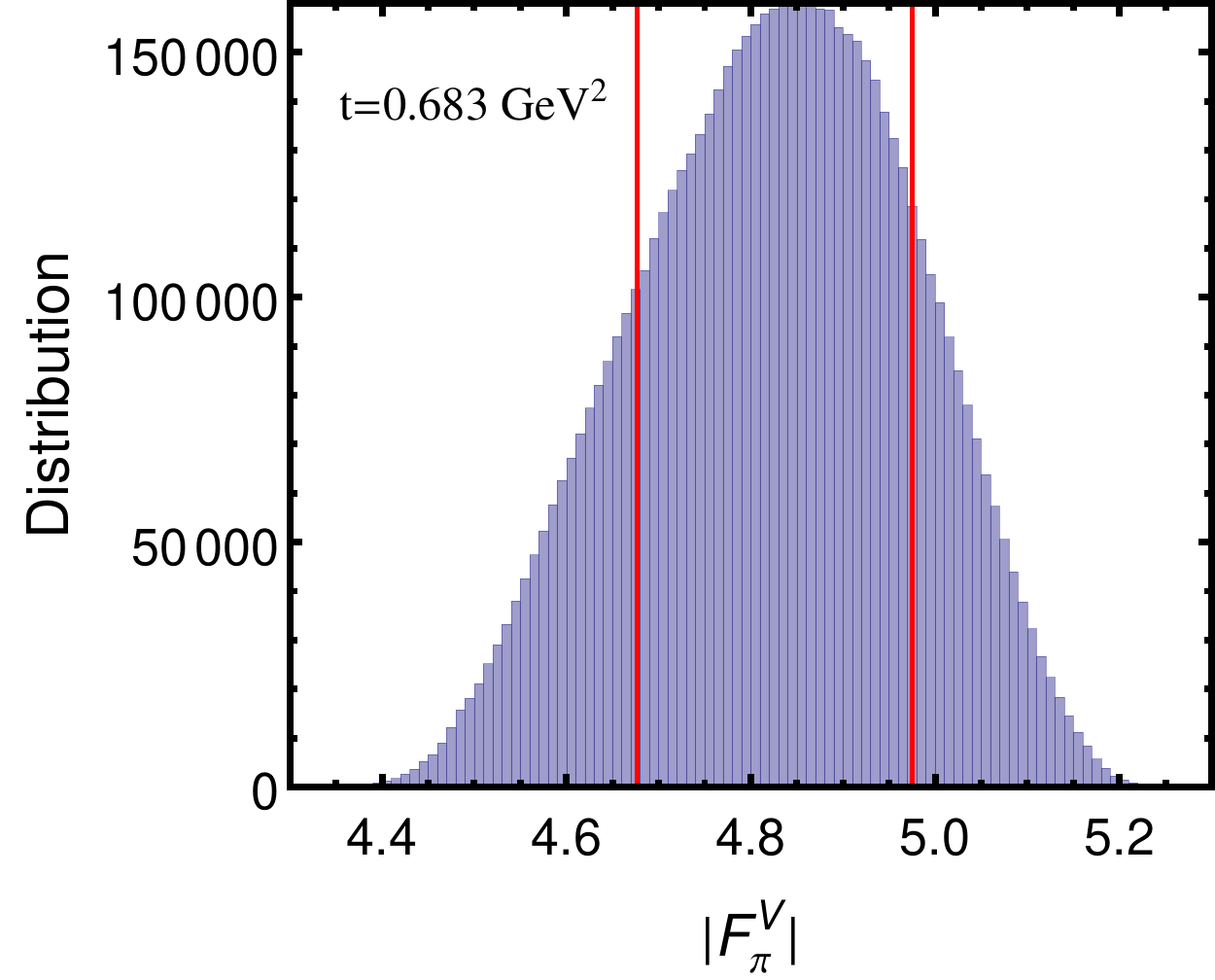}
		\caption{Statistical distribution of $|F_\pi^V(t)|$  at different energies $t$ shown in the legends. To obtain these histogram we use modulus data at  $0.699 \gev$ measured by BABAR, the spacelike point $t_s=-1.6 \gev^2$, and Bern phase. The vertical lines (red) correspond to 68.3\% CL intervals.\label{fig:hist1}}
	\end{center}
	\vspace{0.1cm}
\end{figure*}

\subsection{Methodology of $|F_\pi^V|$ determination}\label{sec:extrap}
We have taken the range $(0.72-0.9) \gev$ as the output region, where we extrapolate the form factor and confront the results with the experimental data.

A nontrivial complication is the fact that the experimental values used as input are known up to statistical and systematic uncertainties. This requires to properly merge the formalism of analytic bounds with statistical simulations. The problem was solved in Refs. \cite{Ananthanarayan:2016mns, Ananthanarayan:2017efc, Ananthanarayan:2018nyx} by generating a large sample of pseudodata, 
achieved by randomly sampling each of the
input quantities with specific distributions based on the measured central values and the quoted errors.
For each point from an input statistical sample, upper and lower bounds on $|F_\pi^V(t)|$ at points $t$ in the output region  have been calculated using the formalism described in the previous section.  Note that the input points had to pass a consistency condition in order to be included in the sample. Indeed,  some of the minors of the determinant ${\cal D}$, mentioned below (\ref{eq:barxi}), involve only input quantities and, if the positivity condition is violated, the corresponding points had to be rejected.  Finally, a set of allowed values  in between the bounds has been uniformly generated, taking into account the fact that all the values between the extreme points are equally valid. The number of generated   points was adapted to the width of the allowed range, being larger fow wider ranges.

In this way, for a specified spacelike and timelike input, we generated a large  sample of output values of $|F_\pi^V(t)|$ at each point $\sqrt{t}$ in range  $(0.72-0.9) \gev$ of interest.  The output distributions turn out to be close to a Gaussian, allowing the extraction of the mean value and the standard deviation (defined as the 68.3\% confidence limit interval). The values obtained with input from each experiment at different energies  and of the different experiments (SND, CMD2, BABAR, KLOE11, KLOE13 and BESIII)  have been then combined using an averaging prescription proposed in \cite{Schmelling:1994pz}, where the correlations between different values are derived from the values themselves. The two spacelike input points have been considered separately and then we took the average of the two predictions for both the central value and the error. The same conservative procedure has been adopted for the phases: we considered the results obtained separately with the Bern and Madrid phases and  also the simple average of the corresponding predictions. The comparison of these results with the experimental values serves as a test of unitarity and analyticity of the experimental determinations in the input and output regions.

\section{Results}\label{sec:res}
We start by illustrating the statistical distributions of the output values of  $|F_\pi^V(t)|$, obtained with a specific input.  
In Figs.  \ref{fig:hist1}  we show these distributions  at several values of $t$, obtained with the pseudodata sample generated with the experimental measurement at $0.699 \gev$ by BABAR,  the first spacelike input (\ref{eq:Huber}) and Bern phase. 
One can note the increasing number of points in the samples with increasing $t$. The explanation is that the allowed bands between the upper and lower bounds become wider at larger energies, and more intermediate points must be generated in order to obtain the correct statistical distribution.
All the distributions are very close to a Gaussian, allowing the extraction of a central value and a standard deviation at 68.3\% confidence level.

From these distributions, by applying the averaging procedure described in Sec. \ref{sec:extrap}, we obtained a central value and a standard deviation for the modulus for all energies in the output region  $(0.72-0.9) \gev$. For convenience, we   present the extrapolated values of  $|F_\pi^V(t)|^2$ at 68.3\% CL in this region as a band denoted as ``allowed band''.  This is the main prediction of our method, which we compare with the experimental values available in the same region.

\begin{figure}[thb]\vspace{0.3cm}
		\includegraphics[width = 7cm]{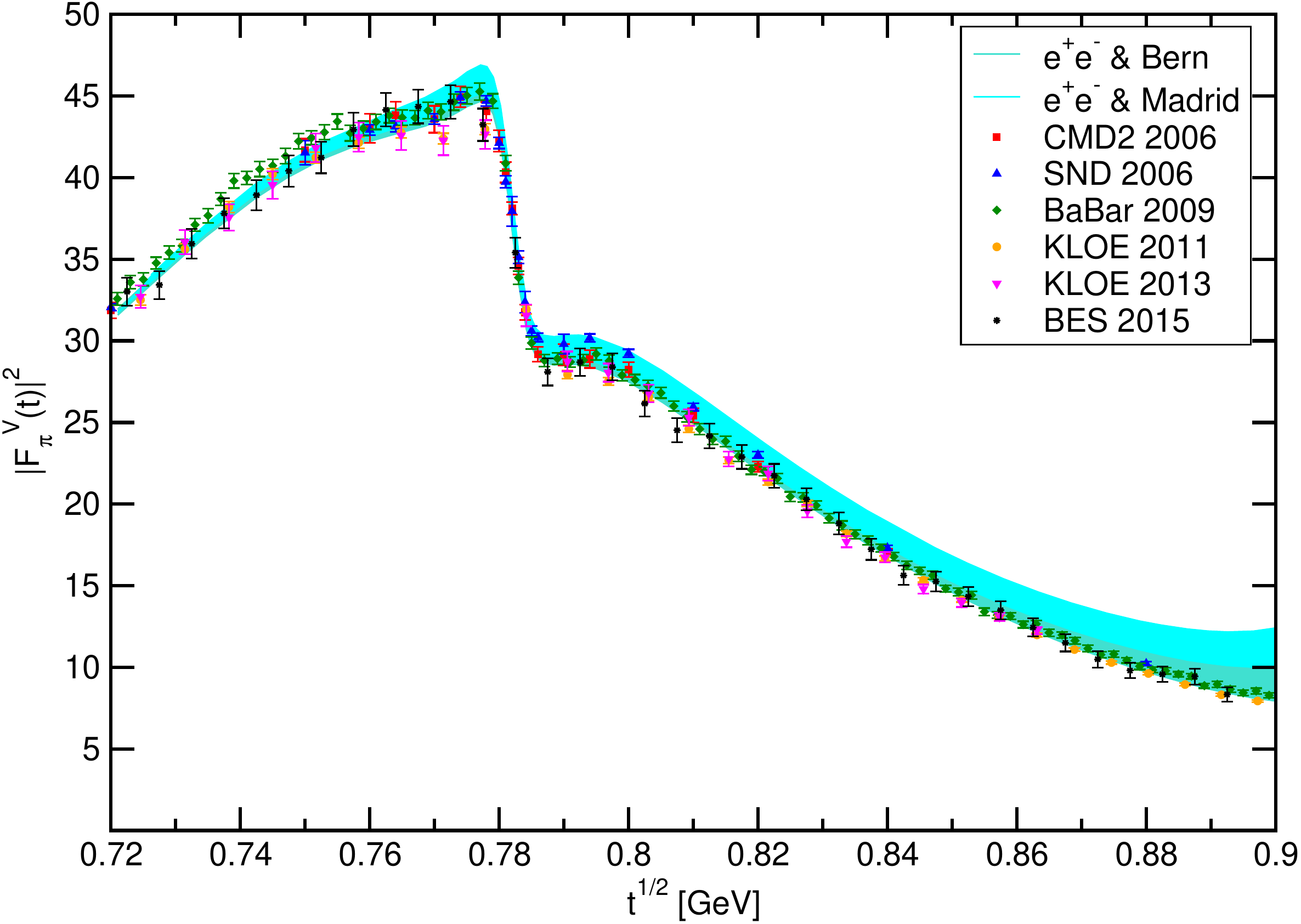}
		\caption{Allowed bands of  $|F_\pi^V(t)|^2$ at 68.3\% CL obtained with input from all $e^+e^-$ experiments. The bands obtained with the  Bern and Madrid  phases are shown separately.\label{fig:epemvsBM}}
	\vspace{0.3cm}
\end{figure}

In Fig. \ref{fig:epemvsBM} we show the allowed bands for   $|F_\pi^V(t)|^2$  in the range $(0.72-0.9) \gev$, obtained by using as input all $e^+e^-$ experiments (SND, CMD2, BABAR, KLOE 11, KLOE 13 and BESIII) in comparison with the  $e^+e^-$  experimental data in the same region. The bands obtained   with the Bern and Madrid phases are shown separately.

Several remarks can be made from this figure. First, below the $\rho$ peak the bands of extrapolated values are rather narrow, competing with the experimental data in this region. Only BABAR and KLOE data are more precise than the prediction. Several BABAR points lie definitely above the allowed band,  while several KLOE points lie below the band. We can say that these points are in certain tension with analyticity and the data from the lower-energy region,  $(0.65-0.71)\gev$. For the other datasets, which  have larger uncertainties, no inconsistencies are seen.

\begin{figure}[thb]\vspace{0.3cm}
		\includegraphics[width = 7cm]{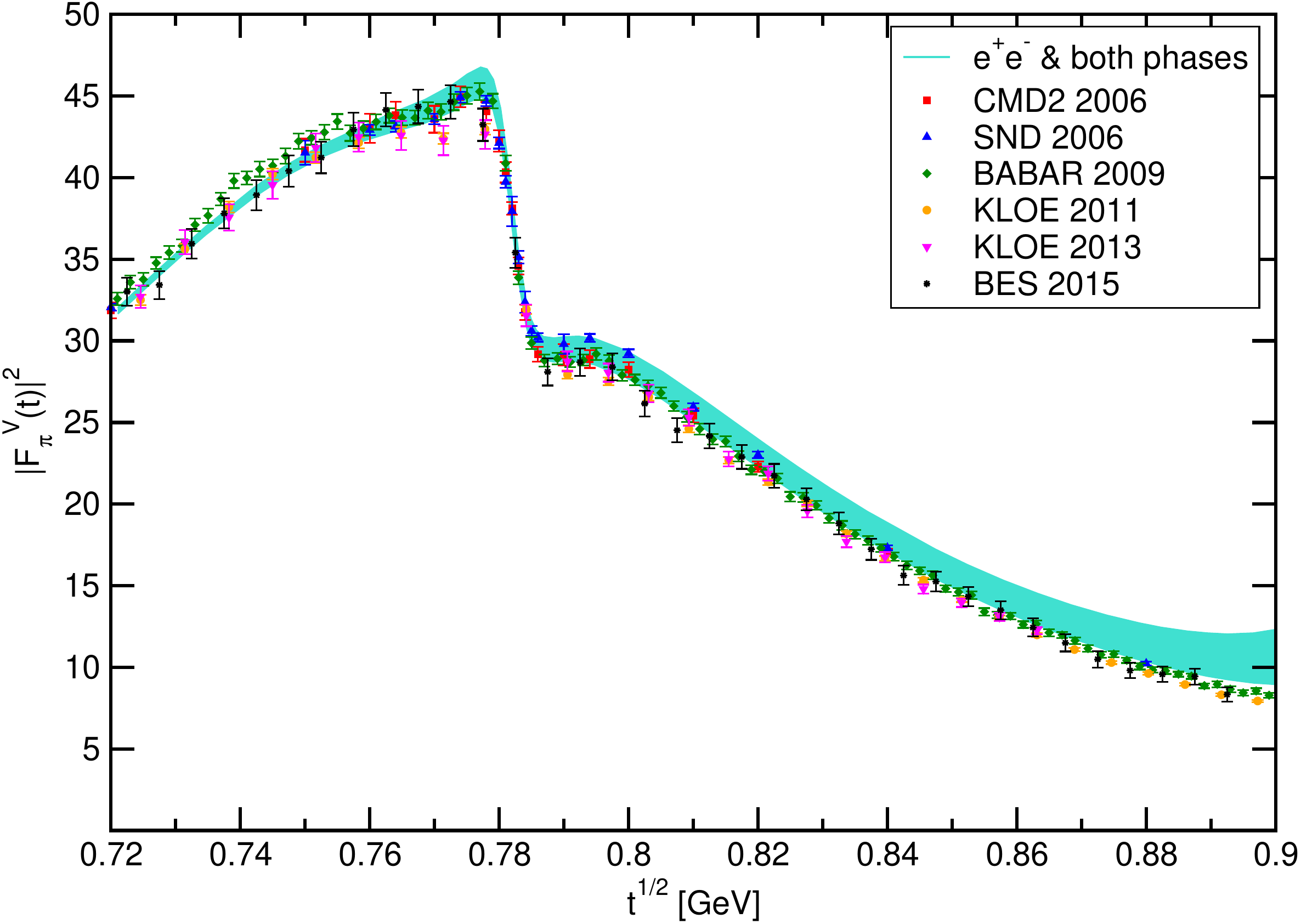}
		\caption{Allowed band obtained with input from all $e^+e^-$ experiments and Bern and Madrid phases.\label{fig:epemBM}}
	\vspace{0.3cm}
\end{figure}

Above the  $\rho$ peak, the allowed bands obtained by  extrapolation become gradually wider, as anticipated, and cannot compete in precision with the experimental data. However, some interesting features can be seen also in this region. All the data lie at the lower edge of the wider band obtained with  the Bern phase \cite{Caprini:2011ky}, and are definitely below  the band obtained with the Madrid phase \cite{ GarciaMartin:2011cn}, which is more narrow since the quoted uncertainties are smaller. Some experimental points near 0.9 GeV are also below the average band obtained with the two phases, presented in Fig.  \ref{fig:epemBM}.

\begin{figure}[thb]\vspace{0.3cm}
		\includegraphics[width = 7cm]{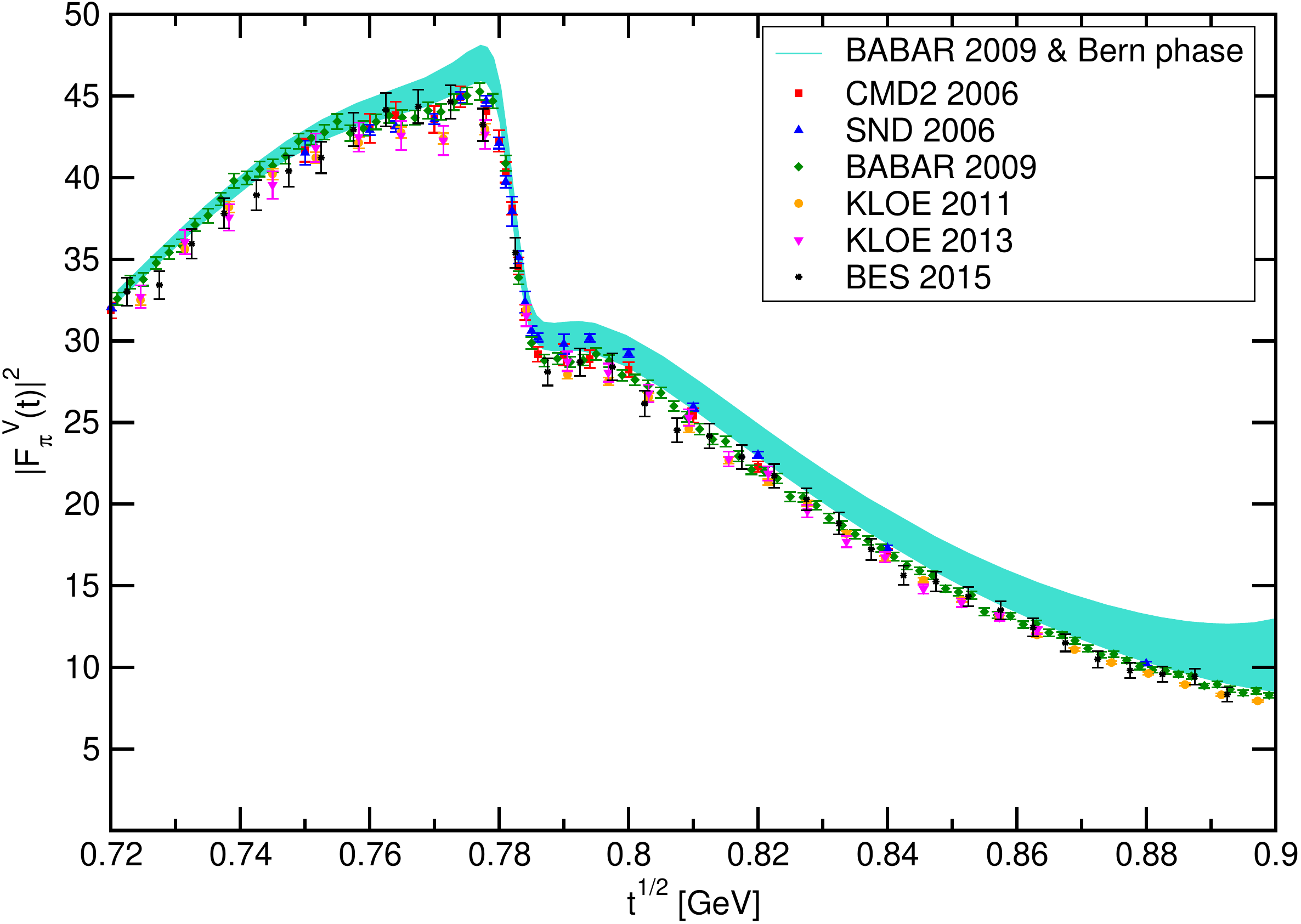}
		\caption{Allowed band using as input only BABAR data and Bern phase.\label{fig:BabarB}}
	\vspace{0.3cm}
\end{figure}

\begin{figure}[thb]\vspace{0.3cm}
		\includegraphics[width = 7cm]{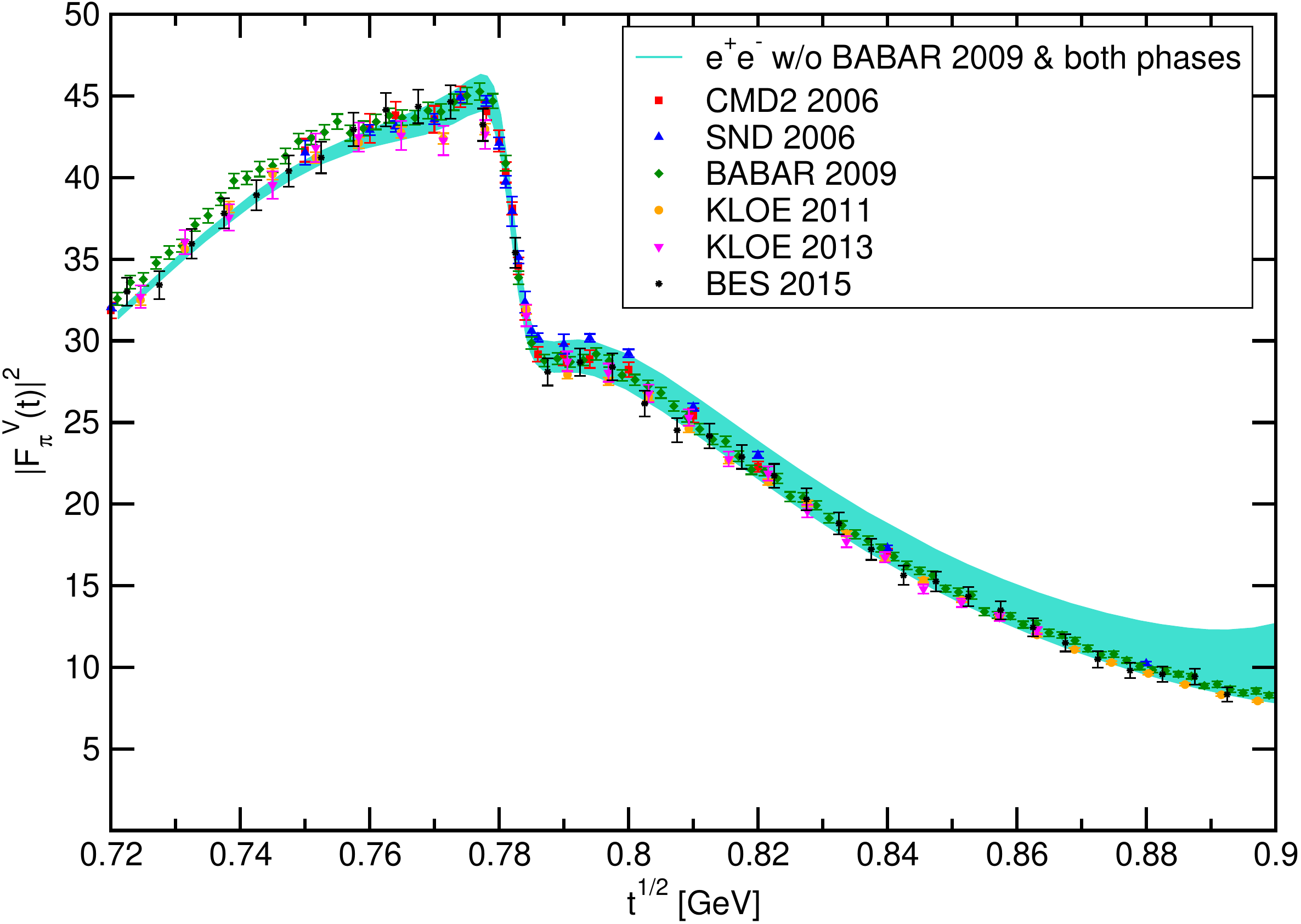}
		\caption{Allowed band using as input $e^+e^-$ data without BABAR, and Bern phase.\label{fig:noBabarB}}
	\vspace{0.3cm}
\end{figure}

The above predictions have been obtained by using as input all the data in the  range  $(0.65-0.71)\gev$. It is of interest to see what happens if we restrict the input to some datasets. In Fig. \ref{fig:BabarB} we show the allowed band obtained using as input only BABAR 2009 data \cite{Aubert:2009ad}  and Bern phase.  One can see that now the BABAR points below the $\rho$ peak are in more agreement with the allowed band, while the KLOE data lie definitely below the band and above the $\rho$ peak all the experimental data are well below the allowed band. As an opposite choice, we show in Fig. \ref{fig:noBabarB} the allowed band obtained using as input all the data except BABAR, with Bern phase used as before. The data are now in more agreement with the allowed band, except the BABAR data below the $\rho$ peak which are definitely higher.

\begin{figure}[thb]\vspace{0.3cm}
		\includegraphics[width = 7cm]{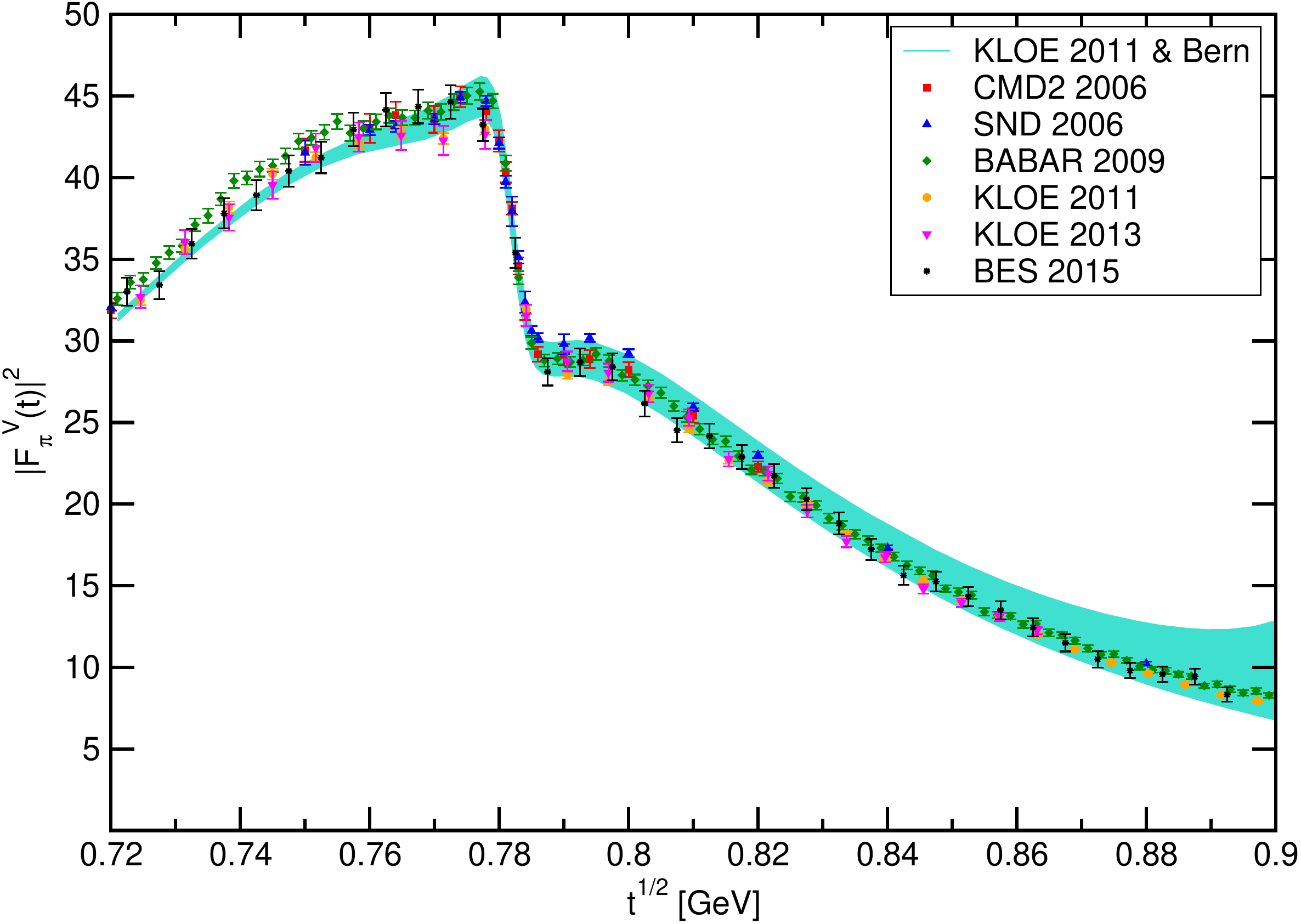}
		\caption{Allowed band using as input only KLOE11 data and Bern phase.\label{fig:Kloe11B}}
	\vspace{0.3cm}
\end{figure}

\begin{figure}[thb]\vspace{0.3cm}
		\includegraphics[width = 7cm]{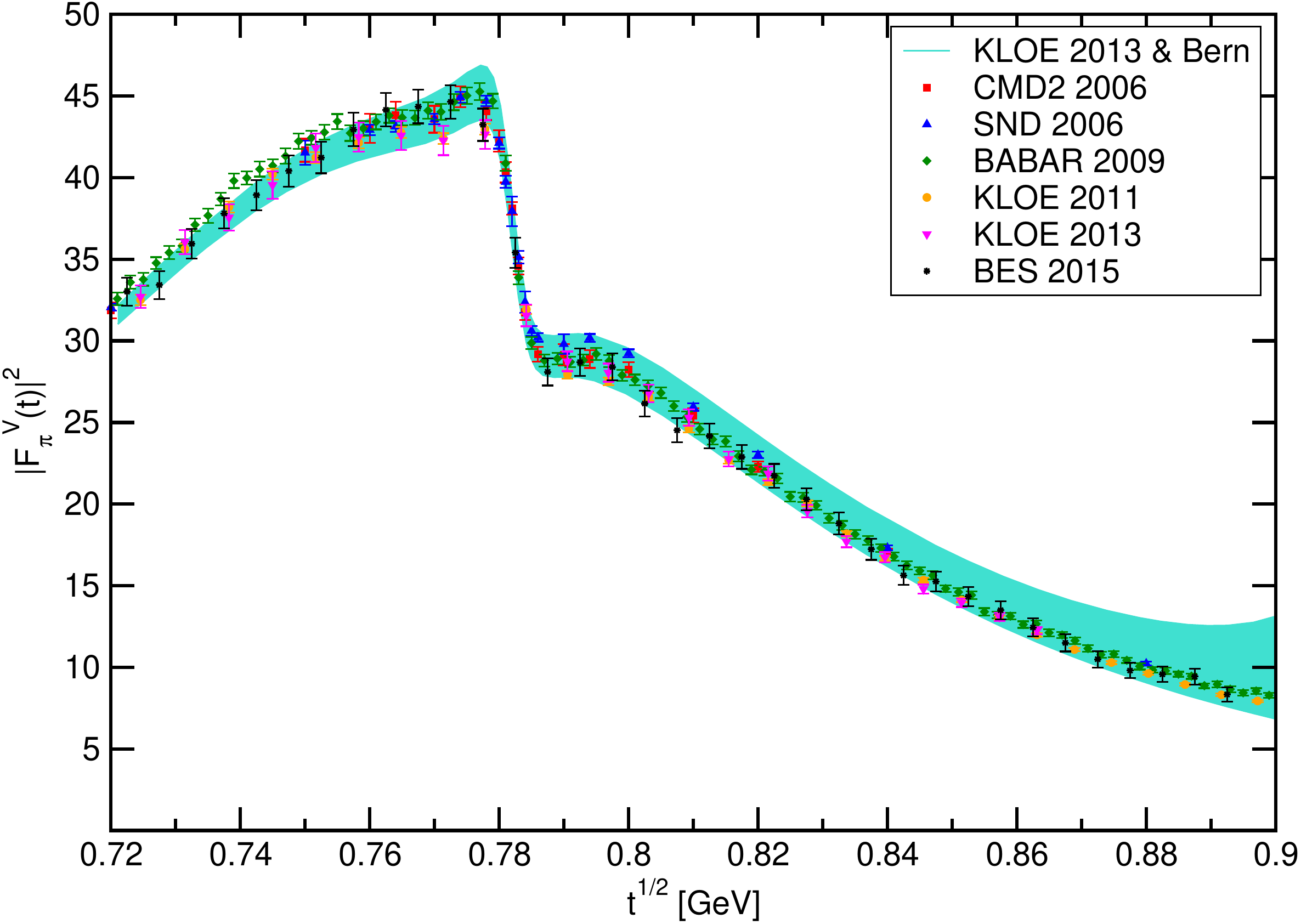}
		\caption{Allowed band using as input only KLOE13 data and Bern phase.\label{fig:Kloe13B}}
	\vspace{0.3cm}
\end{figure}

Finally, in Figs. \ref{fig:Kloe11B} and \ref{fig:Kloe13B} we present the allowed bands obtained using as input only KLOE 2011 \cite{Ambrosino:2010bv} and KLOE 2013 \cite{Babusci:2012rp} data, respectively, and Bern phase.  Now all the experimental points, below and above the $\rho$ peak, are inside the allowed range, except the BABAR data below the  $\rho$ peak, which are clearly above it. We note that the disagreement is more pronounced for the  KLOE 2011 input, and slightly less stringent for the  KLOE 2013 input.

\section{Discussion and conclusions}\label{sec:conc}
In the present paper we have investigated the consistency of the recent high-statistics experimental data on the pion vector form factor with analyticity and unitarity. The aim was to provide additional constraints on the dispersive evaluation of the HVP contribution to the SM value of the muon $g-2$. The work was motivated also by the tensions which exist between the most precise data, BABAR and KLOE, which have not been resolved so far. 

We have used a model-independent formalism, which uses Fermi-Watson theorem (\ref{eq:watson}) in the elastic region and a very conservative integral condition (\ref{eq:L2}) on the modulus in the inelastic region. From these boundary conditions,  rigorous constraints on the values of the form factor at points inside the analyticity domain can be derived, expressed in the particular case of three points as the positivity condition (\ref{eq:posit}) of the determinant (\ref{eq:det}) and of its minors. 

In our analysis, we have exploited this constraint by using as input the experimental data (\ref{eq:Huber}) on the form factor on the spacelike axis and the measurements of the modulus  in the region  $(0.65 - 0.71) \gev$ by the SND, CMD2, BABAR, KLOE 11, KLOE 13 and BESIII experiments. The choice of this region was justified by the fact that data from various experiments are here in better mutual agreement than at higher energies.  It may be noted however that this choice is an educated guess and cannot necessarily be considered very rigorous. In particular, some tensions between BABAR and KLOE data exist even in this region.

Using a specific input,  we have derived upper and lower bounds on the modulus in the higher-energy region  $(0.72-0.9)\gev$, which includes the $\rho$ peak. The bounds derived from analyticity have been converted into central values and standard deviations at 68.2\% CL by using numerical simulations on pseudodata samples, and have been presented as a band of ``allowed values''. The aim was to compare the bands obtained by analytic extrapolation with the experimental data available in the same region, where the tensions between  BABAR and KLOE are larger. 

We emphasize that in the present formalism the analytic extrapolation  is performed without  a specific parametrization of the form factor, like the Gounaris-Sakurai formula or the Omn\`es-like parametrizations used in the literature.  Moreover, the results are optimal for a definite input and are rigorously independent on the unknown phase of the form factor above the inelastic threshold. For completeness, we gave in the Appendix \ref{sec:A} formal argument and a detailed proof of this property, presented here for the first time.   

The results presented in Sec. \ref{sec:res} indicate some inconsistencies between the data and the extrapolated band. The tension between BABAR and KLOE  is manifest and exhibits new facets. As shown in Fig. \ref{fig:BabarB}, using as input only BABAR data in the region  $(0.65 - 0.71)\gev$,  the KLOE data below the $\rho$ peak are definitely lower than the extrapolated band. More impressively,  above the $\rho$ peak all the data, including BABAR, lie below the allowed band. If, on the other hand, only KLOE data in the region  $(0.65 - 0.71)\gev$ are used as input, Figs. \ref{fig:Kloe11B} and   \ref{fig:Kloe13B}  show that all the data above 0.72 GeV are consistent with the allowed band, except for BABAR data below the $\rho$ peak, which lie definitely above the band. 

A somewhat surprising feature is that all the data  above the $\rho$ peak  appear to be situated at the lower edge of the extrapolated  band. This feature is more dramatic when only BABAR data in the region  $(0.65 - 0.71) \gev$  are used as input, as seen in Fig. \ref{fig:BabarB}, but is obtained also with input from all experiments, or from all experiments except BABAR,  as seen in Figs. \ref{fig:epemBM} and \ref{fig:noBabarB}, respectively. 

It is difficult at present to find a simple explanation for this result. Thus, it is unlikely that the resolution of the puzzle lies in insufficient knowledge of the phase.  An important ingredient in our approach is the phase shift $\delta_1^1$, but this function is calculated with precision in $\pi\pi$ dispersion theory. Moreover, we used two phases, calculated independently in  \cite{Caprini:2011ky} and  \cite{GarciaMartin:2011cn}, and the results are consistent, with a somewhat wider bands obtained with Bern phase, which has larger uncertainties near 0.9 GeV. 

A global correction factor, with large values precisely in the region of interest, might offer a possible explanation. We recall in this context the so-called $\rho-\gamma$ mixing factor, calculated in \cite{Jegerlehner:2011ti}, whose effect is to slightly push  upward the modulus of the form factor extracted from $e^+e^-$ data above the $\rho$ peak. 
However, for the moment we refrain from further speculating on possible explanations of the fact that the data from the region $(0.65 - 0.71) \gev$ seem to require higher values than the experimental measurements of the modulus  above the $\rho$ peak, especially above 0.8 GeV.

Instead, we briefly comment on the implications of this result on the HVP  contribution to the muon $g-2$. As it is known, the  two-pion LO contribution to $a_\mu$, which does not contain the vacuum polarization effects but includes one-photon final-state radiation (FSR),
is expressed as an integral over the modulus squared on the pion form factor (for the explicit formulae see for instance Sec. II of \cite{Ananthanarayan:2016mns}).

Using the central value and the error of the extrapolated band shown in Fig. \ref{fig:epemBM}, we obtained for the contribution from the region $[0.8,\, 0.9] \gev$ the value
$a_\mu[0.8,\,0.9]=(70\pm 4)\times 10^{-10}$. On the other hand, the values obtained for the same quantity from the direct integration of the data are (in units of $10^{-10}$) $67.5(4)(6)$ \cite{Davier:2019can} and $66.6(3)$ \cite{Keshavarzi:2019abf}, while the fit of the form factor performed in \cite{Colangelo:2018mtw}  gives $66.6(4)$. 
The value obtained by us from analytic extrapolation  is much less precise than  these estimates, as could be anticipated actually from the figures, and is consistent with them within the uncertainties. However, the fact that the central value is higher by about 3.8 units of  $10^{-10}$ is a intriguing and deserves further investigations.

To summarize, in this work we have carried out the logical extension of our previous work which combines the method of unitarity bounds with detailed Monte Carlo simulations to yield determinations of the radius and the values of the form factor below 0.63 GeV, to the region above 0.72 GeV in the present work (as the timelike input data are in the region $0.65-0.71\gev$) which is essentially around the $\rho$.  Since the bounds are now weaker we do not have a high precision determination but rather a detailed consistency test of the compatibility of the data in this region with unitarity and analyticity constraints.  Our work reveals some puzzles as discussed in detail in the foregoing.  It is a detailed merger of theoretical methods and experimental data in one of the few systems in the low-energy strong interaction sector where high quality data is available in several kinematic regimes.

\subsection*{Acknowledgments}  I.C. acknowledges support from the   Romanian Ministry of Education and Research, Contract PN 19060101/2019-2022. D.D. is supported by the DST, Government of India, under INSPIRE Fellowship (No. DST/INSPIRE/04/2016/002620).

\appendix

\section{Independence of the bounds on the phase $\phi(t)$ for $t>\tin$}\label{sec:A}

The Omn\`es function $\omnes(t)$ defined in (\ref{eq:omnes}) is not unique, 
as it involves the arbitrary function $\phi(t)$ for $t>\tin$. 
However, as argued in  \cite{Caprini:1999ws, Abbas:2010jc},  a change
of the function  $\phi(t)$ for $t>\tin$ is equivalent with a multiplication by a function analytic and without zeros  in $|z|<1$ (i.e., a so-called outer
function). According to the general theory of analytic functions of Hardy class \cite{Duren}, the
multiplication by an outer function leads to an equivalent class of analytic functions. Therefore, the solution of the extremal problems defined on this class does not change, which means that the bounds do not depend on the phase  $\phi(t)$ for $t>\tin$. 

 It is instructive to give also an explicit proof of this independence. We first note that the functions $w(z)$ are $g(z)$ do not involve the arbitrariness in question: $w(z)$ is given explicitly in (\ref{eq:outer}), and $g(z)$ is analytic in $|z|<1$ and subject only to the condition (\ref{eq:gI1}).  We must consider therefore only the functions $\omnes(t)$ and $\omega(t)$. More precisely, as seen from (\ref{eq:gFn}) and (\ref{eq:gFn1}), we are interested in the ratio of their values  for $t<\tin$. 

It is convenient to write    $\omnes(t)$ as
\beq\label{eq:omnesprod}
\omnes(t)= \omnes_1(t) \cdot \omnes_2(t),
\eeq
where
\bea\label{eq:omnes12}
&& \omnes_1(t)=\exp \left(\D\frac {t} {\pi} \int^{\tin}_{t_+} dt' 
\D\frac{\delta_1^1(t^\prime)} {t^\prime (t^\prime -t)}\right),\nonumber\\
&&\omnes_2(t)=\exp \left(\D\frac {t} {\pi} \int^{\infty}_{\tin} dt' 
\D\frac{\phi(t^\prime)} {t^\prime (t^\prime -t)}\right).
\eea
From (\ref{eq:omega}) and (\ref{eq:modomnes}) it follows that we can write also
\beq\label{eq:omegaprod}
\omega(t)= \omega_1(t)\cdot \omega_2(t)
\eeq
where
\bea\label{eq:omega12}
\omega_1(t)&=& \exp \left(\frac {\sqrt{\tin - t}}{\pi} \int^{\infty}_{\tin}\frac{d t'}  {\sqrt{t' - \tin} (t' - t)}\right.\nonumber\\
&\times& \left.\frac{t'}{\pi} \int^{\tin}_{t_+} \frac{\delta_1^1(t'') dt''} {t'' (t'' -t')}, \right), \nonumber\\
\omega_2(t)&=& \exp \left(\frac {\sqrt{\tin - t}}{\pi} \int^{\infty}_{\tin}\frac{d t'}  {\sqrt{t' - \tin} (t' - t)}\right.\nonumber\\
&\times& \left.\frac{t'}{\pi}\text{PV} \int^{\infty}_{\tin} \frac{\phi(t'') dt''} {t'' (t'' -t')} \right).
\eea
Here we indicated that the PV has to be taken only in $\omega_2(t)$, where the integration variables $t''$ and $t'$ can coincide.

The functions  $\omnes_1(t)$ and $\omega_1(t)$ depend only on the known phase $\delta_1^1(t)$ in the elastic region $t<\tin$. We must investigate therefore only the ratio of  $\omnes_2(t)$ and $\omega_2(t)$, which   separately depend on the arbitrary phase $\phi(t)$ for $t>\tin$. We recall that we have to evaluate these functions for $t<\tin$. Moreover, we note that the Principal Value is equal actually to the real part of the corresponding integral, as follows from the formal Plemelj relation
\beq
\frac{1}{x-z-i \epsilon}=\text{PV}\frac{1}{x-z}+ i \pi \delta(z-x),
\eeq 
Using the identity
\beq
\frac{t'}{t''(t''-t')}=\frac{1}{t''-t'}-\frac{1}{t''}
\eeq
and permuting the order of integration, we write $\omega_2(t)$ as:

\begin{widetext}
	\beq\label{eq:omega2}
	\omega_2(t)= \exp \left(\frac {\sqrt{\tin - t}}{\pi}\, \left[\int^{\infty}_{\tin} \phi(t'') dt'' \frac{1}{\pi}\text{Re} \int^{\infty}_{\tin}\frac{d t^\prime}  {\sqrt{t' - \tin} (t' - t) (t''-t')}   -
	\int^{\infty}_{\tin} \frac{\phi(t'') dt''}{t''}\frac{1}{\pi} \text{Re} \int^{\infty}_{\tin}\frac{ d t'}  {\sqrt{t' - \tin} (t' - t)} \right]\right).
	\eeq
\end{widetext}

It is convenient to use in the first integral the identity
\beq
\frac{1}{((t'-t)(t''-t')}=\left[\frac{1}{t''-t'}+\frac{1}{t'-t}\right]\,\frac{1}{t''-t}.
\eeq
Then the integrals upon $t'$  in (\ref{eq:omega2}) can be performed exactly using the generic dispersion relation
\beq
\frac{1}{\pi} \int^{\infty}_{\tin} \frac{dt'}{\sqrt{t'-\tin}(t'-y)}=\frac{1}{\sqrt{\tin-y}}
\eeq
satisfied by the function $1/\sqrt{\tin-y}$, which is analytic of real type in the $y$ complex plane cut for $y>\tin$  and has an imaginary part equal to $1/\sqrt{t'-\tin}$  on the upper edge of the cut.  

In (\ref{eq:omega2}), $y$ is replaced either by $t''$, when the result $1/\sqrt{\tin-t''}$ is purely imaginary and does not contribute to the real part, or by $t$, when the result  $1/\sqrt{\tin-t}$ is real and can be simplified with the factor $\sqrt{\tin - t}$ appearing in the numerator. Then it is easy to see that the two terms, with denominators $t''-t$ and $t''$, combine into a single term, leading to 
\beq\label{eq:oomega2fin}
\omega_2(t)=\exp \left(\D\frac {t} {\pi} \int^{\infty}_{\tin} dt'' 
\D\frac{\phi(t'')} {t'' (t'' -t)}\right).
\eeq
But this term coincides with $\omnes_2(t)$ defined in (\ref{eq:omnes12}), so that we obtain 
\beq
\frac{\omega(t)}{\omnes(t)}=\frac{\omega_1(t)}{\omnes_1(t)},\quad\quad t<\tin.
\eeq 

From this equality it follows that the relations (\ref{eq:gFn}) and (\ref{eq:gFn1}) contain only the quantities $\omega_1(t_n)$ and $\omnes_1(t_n)$, independent on the phase $\phi(t)$ for $t>\tin$, which proves the independence on this phase of the bounds derived from (\ref{eq:posit}) and (\ref{eq:det}). 

The rigorous independence of the bounds on the phase  $\phi(t)$ above $\tin$ was checked numerically  with various choices of this function. In particular,  one can take $\phi(t)=0$ for $t>\tin$, which implies $\omnes_2(t)=1$ and $\omega_2(t)=1$. In this case, the phase in the integral (\ref{eq:omnes})  has a discontinuity at $t=\tin$, which  leads to a logarithmic singularity of $\ln|\omnes_1(t)|$ at this point. This implies an end singularity of the form $1/\sqrt{t'-\tin}\, \ln(t'-\tin)$ in the integrand of (\ref{eq:omega}) at $t=\tin$. However, the singularity is integrable and can be handled numerically with precision.


\end{document}